\documentclass[reprint,superscriptaddress,amsmath,amssymb,aps,prr,twocolumn,groupedaddress]{revtex4-2}
\usepackage[utf8]{inputenc}
\usepackage[dvipsnames]{xcolor}
\usepackage{graphicx, physics, dsfont}
\usepackage[normalem]{ulem}
\allowdisplaybreaks
\def\ii{{\rm i}}  
\def\rb{{\bf r}}
\def\kb{{\bf k}}
\def\gb{{\bf g}}
\def\Ab{{\bf A}}
\def\Db{{\bf D}}
\def\Pb{{\bf P}}
\def\Gab{{\bf \Gamma}}
\def\jop{\hat{\mathcal{O}}}
\def\db{\boldsymbol{\wp}}
\def\dcrit{d_{\mathrm{critical}}}
\def\dk{\mathrm{d}k_z}
\def\dkb{\mathrm{d}\kb}
\def\sge{\hat{\sigma}_{ge}}

\def\bra#1{\mathinner{\langle{#1}|}}
\def\ket#1{\mathinner{|{#1}\rangle}}

\begin{document}
\title{Dicke superradiance in ordered lattices: dimensionality matters}

\author{Eric Sierra}
\author{Stuart J. Masson}
\author{Ana Asenjo-Garcia}
\email{ana.asenjo@columbia.edu}
\affiliation{Department of Physics, Columbia University, New York, NY 10027, USA}
\date{\today}

\begin{abstract}
Dicke superradiance in ordered atomic arrays is a phenomenon where atomic synchronization gives rise to a burst in photon emission. This superradiant burst only occurs if there is one -- or just a few -- dominant decay channels. For a fixed atom number, this happens only below a critical interatomic distance. Here we show that array dimensionality is the determinant factor that drives superradiance. In 2D and 3D arrays, superradiance occurs due to constructive interference, which grows stronger with atom number. This leads to a critical distance that scales sublogarithmically with atom number in 2D, and as a power law in 3D. In 1D arrays, superradiance occurs due to destructive interference that effectively switches off certain decay channels, yielding a critical distance that saturates with atom number. Our results provide a guide to explore many-body decay in state-of-the art experimental setups.
\end{abstract}

\maketitle

\section{Introduction}

\begin{figure*}
    \centering
    \includegraphics[width=0.95\textwidth]{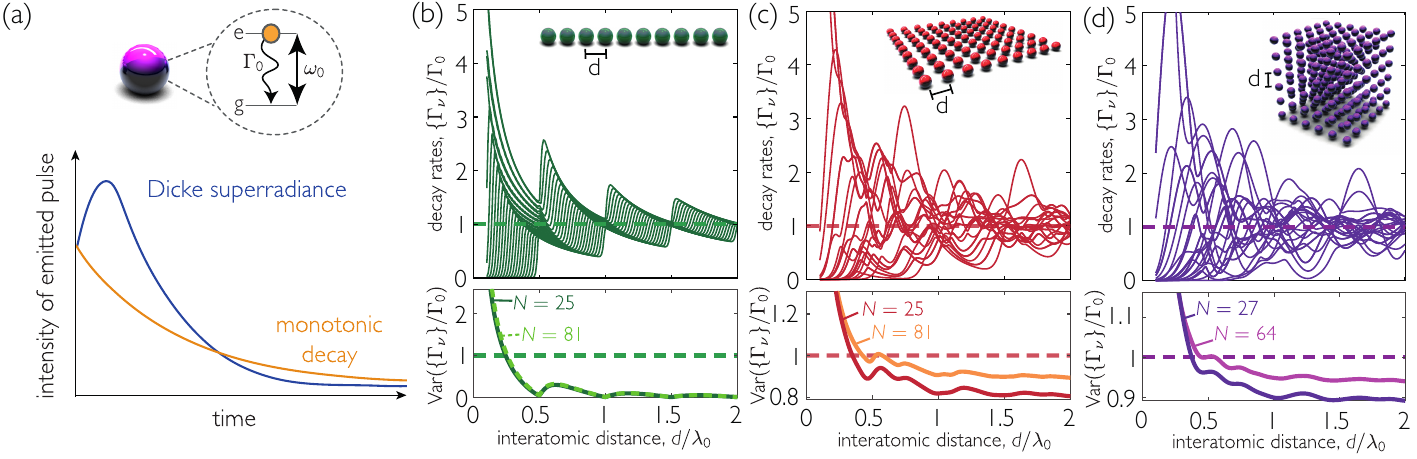}
    \caption{(a) Schematic of Dicke superradiance, which results from atomic correlations and leads to a burst in the emitted intensity, in contrast to the monotonic decay from uncorrelated atoms. In ordered arrays, the interatomic distance controls the crossover between the two regimes. Atoms are modeled as two-level systems with transition frequency $\omega_0$ between excited and ground states and spontaneous emission rate $\Gamma_0$. (b-d) Collective decay rates (top) and their variance (bottom) as a function of interatomic distance $d$ for atomic arrays of different dimensions. (b) $N=25$ atoms form a 1D array, and the transition polarization is perpendicular to the chain. (c) $N=5^2=25$ atoms form a 2D lattice with polarization axis out-of-plane. (d) $N=3^3=27$ atoms are arranged in a 3D lattice, with polarization axis aligned with one of the main axes of the array.}
    \label{Figure1}
\end{figure*}

Collective emission has been a fundamental problem in quantum optics since the notion was introduced by Dicke in 1954~\cite{Dicke54}. Dicke realized that nearby atoms must interact via shared electromagnetic field modes, fundamentally altering their optical properties. He considered the case of a fully excited ensemble of emitters located at a single point. In stark contrast to the exponentially-decaying pulse emitted by independent (i.e., far-separated) entities, emitters at a point synchronize, locking in phase as they decay, and emitting a short burst of light that initially rises in intensity~\cite{Dicke54,Rehler71,Gross82,BenedictBook} [see Fig.~\ref{Figure1}(a)]. This ``superradiant burst'' has become a hallmark of collective phenomena in quantum optics and has been observed in a variety of physical systems~\cite{Skribanowitz73,Raimond82,Inouye99,Scheibner07,Slama07,Raino18,Ferioli21PRL}. Generally, experiments are performed in dense disordered systems, where interparticle separations can be very small, or in a cavity, where the restriction of the field to a single confined mode emulates the condition of atoms at a point.

In extended ordered arrays in free space, the geometry and dimensionality of the lattice define the atomic decay properties due to position-dependent dipole-dipole interactions. For example, in ordered arrays with subwavelength interatomic separation, subradiant states with an extremely enhanced lifetime emerge~\cite{Mewton07,Zoubi10,Sutherland16}. These can be used to guide light as ``atomic waveguides''~\cite{Chui15,Asenjo17PRX,Needham19} or ``atomic dielectrics''~\cite{Masson20PRR,Patti21,Brechtelsbauer21,CastellsGraells21}, and to improve the fidelity of protocols for quantum information storage~\cite{Facchinetti16,Asenjo17PRX,Manzoni18} and  metrology~\cite{Kramer16,Henriet19}, among other applications. Arranging single atoms in ordered patterns has become an experimental reality~\cite{Bakr10,Sherson10,Kim16,Endres16,Barredo16,Kumar18,Norcia18,Saskin19,OhlDeMello19}. Interatomic separations in these platforms can be small enough that the optical response is strongly modified by photon interference effects. Predictions that a two-dimensional array acts as an atomically-thin mirror have been demonstrated~\cite{Bettles16PRL,Shahmoon17,Rui20}, and site-dependent frequency shifts due to dipole-dipole interactions have been measured~\cite{Glicenstein20}. 

The role of geometry in many-body (i.e., many-photon, or ``Dicke'') superradiance has not yet been completely elucidated. The introduction of spatially-varying dipole-dipole interactions to Dicke's original model causes position-dependent frequency shifts that lead to dephasing, damping superradiance~\cite{Friedberg72,Coffey78}. This is significant in disordered systems, and points to the key role of geometry, as it determines the spatial pattern of these frequency shifts~\cite{Stroud72,Friedberg74OptComm,Banfi75}. However, dephasing is reduced in large ordered arrays, as the frequency shifts are predominantly homogeneous~\cite{Coffey78,Gross82,BenedictBook}. In such systems, the primary source of dephasing is competition between multiple decay channels~\cite{Masson20PRL,Masson22}.

Here, we study the onset of superradiance in lattices of different dimensionalities. To do so, we harness a technique derived in our previous work~\cite{Masson22}. It enables us to deduce the minimal conditions for Dicke superradiance without calculating the full dynamical evolution of the system, simply by analyzing the statistics of the first two emitted photons~\cite{Masson22}. We analytically and numerically demonstrate that the microscopic origin of superradiance is highly dependent on the array dimensionality. For two-dimensional (2D) and three-dimensional (3D) arrays, it occurs because of constructive interference. In one-dimensional (1D) arrays, it is due to destructive interference. In contrast to Dicke's original work, which assumes that all atoms are confined to a volume of dimensions much smaller than the transition wavelength, we show that superradiance survives in (high-dimensional) arrays where the \textit{smallest} interparticle distance is larger than a wavelength. The distance increases with atom number, sublogarithmically in 2D and as a power law in 3D. In contrast, it saturates to a certain bound in 1D arrays.

\section{Methods}

We consider ordered arrays of $N$ two-level atoms of resonance frequency $\omega_0$, resonance wavelength $\lambda_0=2\pi c/\omega_0$, and spontaneous emission rate $\Gamma_0$ 
[see Fig.~\ref{Figure1} (a)] arranged at positions $\{\rb_i\}$ with $d$ the smallest interatomic distance in the set $\{|\rb_i-\rb_j|\}$. The atoms interact via the electromagnetic field, which is traced out using a Born-Markov approximation~\cite{Gruner96,Dung02}. The atomic density matrix $\rho$, evolves according to the master equation
\begin{equation}\label{masterequation}
    \dot{\rho}=-\frac{i}{\hbar}\left[\mathcal{H},\rho\right]+\sum_{\nu=1}^N \frac{\Gamma_{\nu}}{2}\left(2\hat{\mathcal{O}}_{\nu}\rho\hat{\mathcal{O}}_{\nu}^{\dagger}-\rho\hat{\mathcal{O}}_{\nu}^{\dagger}\hat{\mathcal{O}}_{\nu} -\hat{\mathcal{O}}_{\nu}^{\dagger}\hat{\mathcal{O}}_{\nu}\rho\right).
\end{equation}
In the above expression, the Hamiltonian is given by
\begin{equation}
    \mathcal{H} = \hbar \sum\limits_{i=1}^N \omega_0\hat{\sigma}_{ee}^i + \hbar \sum\limits_{i,j=1}^N J^{ij}\hat{\sigma}_{eg}^i\hat{\sigma}_{ge}^j,
\end{equation}
where $\hat{\sigma}_{ge}^i = \ket{g_i} \bra{e_i}$ is the atomic lowering operator for atom $i$, with $\ket{g_i}$ and $\ket{e_i}$ the atomic ground and excited states, respectively. The Lindblad operator has been diagonalized into the action of $N$ collective jump operators, $\{\mathcal{\jop_\nu}\}$ with decay rates $\{\Gamma_\nu\}$~\cite{Carmichael00,Clemens03}, found as the eigenstates and eigenvalues of the dissipative interaction matrix $\mathbf{\Gamma}$ with elements $\Gamma^{ij}$. The coherent and dissipative interaction rates read
\begin{equation}
    J^{ij}-i\frac{\Gamma^{ij}}{2}=-\frac{\mu_0\omega_0^2}{\hbar}\db^*\cdot \mathbf{G}_0(\mathbf{r}_i,\mathbf{r}_j,\omega_0)\cdot\db, \label{eq:green}
\end{equation}
where $\db$ is the dipole matrix element of the atomic transition and $\mathbf{G}_0(\mathbf{r}_i,\mathbf{r}_j,\omega_0)$ is the propagator of the electromagnetic field between points $\rb_i$ and $\rb_j$. Each jump operator represents the emission of a photon into a particular decay channel (with some specific far-field profile). Jump operators act on all atoms, and can be expressed as superpositions of atomic lowering operators, $\jop_\nu = \sum_{i=1}^N \alpha_{\nu,i} \hat{\sigma}_{ge}^i$, where $\alpha_{\nu,i}$ represents some spatial profile over the ensemble.

We define the minimal condition for Dicke superradiance to have the first photon enhance the subsequent one or, equivalently, to have a greater than unity  second-order correlation function at $t=0$ (following our previous work~\cite{Masson22}). The correlation function can be calculated exactly for an initially fully-inverted state and yields a condition on the variance of the set of decay rates~\cite{Masson22}, i.e.,
\begin{align}
\text{Var}\left(\frac{\left\{\Gamma_{\nu}\right\}}{\Gamma_0}\right)\equiv \frac{1}{N}\sum_{\nu =1}^{N} \left(\frac{\Gamma_{\nu}^2}{\Gamma_0^2}-1\right)>1.
\end{align}

The problem of identifying the minimal conditions for a superradiant burst is thus reduced to finding the eigenvalues of the dissipative interaction matrix, an operation that scales polynomially with atom number. Taking advantage of the symmetries of the problem, we calculate the variance of the decay rates without diagonalizing the matrix, requiring only $O(N)$ steps [see Appendix A]. This allows us to study arrays of $N\simeq 10^7$ atoms.
 
For infinite ordered arrays, we calculate the variance by making the prescription
\begin{equation}
\sum\limits_{\nu=1}^N \rightarrow N\left( \frac{d}{2\pi} \right)^n \int \dkb,
\end{equation}
where $n$ is the number of dimensions of the array and $\kb$ is an $n$-dimensional vector. We find the condition for a superradiant burst in the limit $N\rightarrow\infty$ to be:
\begin{equation}\label{condition}
\frac{1}{2}\left(\frac{d}{2\pi}\right)^n\int  \left( \frac{\Gamma(\kb)}{\Gamma_0}\right)^2\;\dkb > 1.
\end{equation}

\color{black}

 As $d$ increases, the decay rates cluster around the single-atom decay rate $\Gamma_0$, decreasing the variance. Figures~\ref{Figure1}(b-d) show the decay rates as a function of interatomic distance $d$, for all array dimensionalities showing ``revivals'' or ``geometric resonances'' at particular distances due to long-range interactions arising from $1/r$ terms in the Green's tensor~\cite{Nienhuis87,Bettles16PRA}, which prevent the variance from decreasing monotonically with $d$. This means that establishing regions of superradiance is non-trivial. As $d$ is increased, superradiance can be lost, but then re-established by a geometric resonance~\cite{Masson22}. Here, we define $\dcrit$ as the distance at which $\text{Var}\left(\left\{\Gamma_{\nu}\right\}/\Gamma_0\right)=1$ (there can thus be multiple values for $\dcrit$ for a given array).

\section{1D arrays}

\begin{figure}[b]
    \centering
    \includegraphics[width=0.475\textwidth]{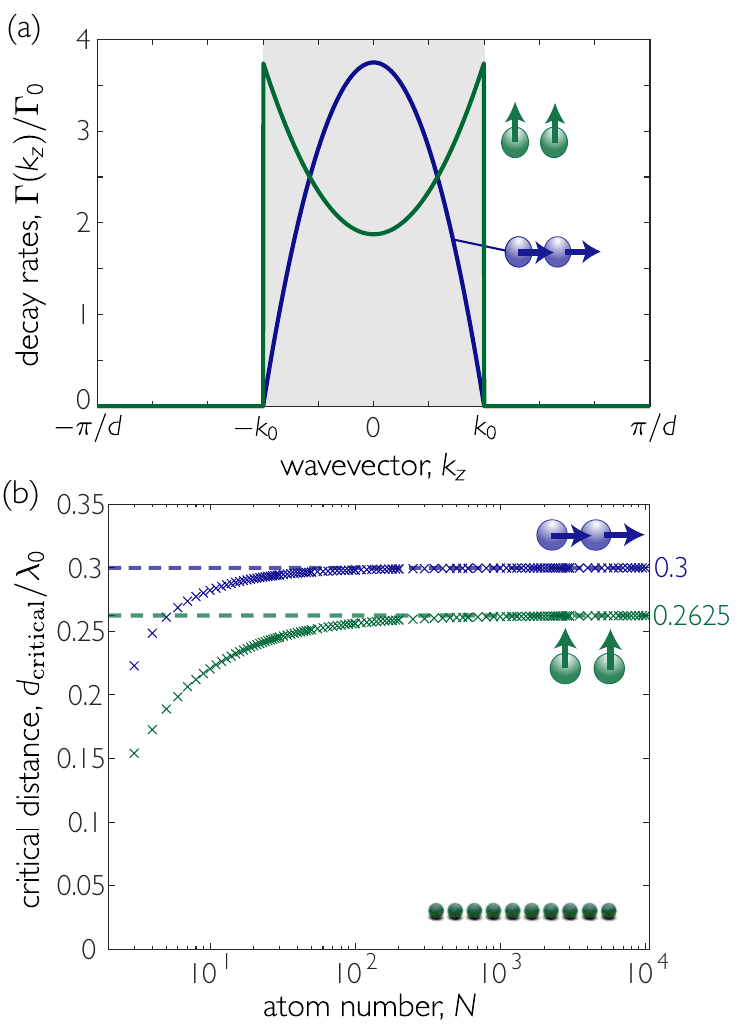}
    \caption{The critical distance (beyond which there is no Dicke superradiance) saturates with atom number in 1D. (a) Decay rates in the first Brillouin zone, for both polarizations (parallel, in blue; and perpendicular, in green) for an infinite 1D array of $d=0.2\lambda_0$. The shaded region represents the light cone, where modes are radiative. (b) Scaling of the critical distance with atom number. Dashed lines show the analytical result for $\dcrit$ obtained for infinite arrays [see Eq.~\eqref{1Dcrit}].}
    \label{Figure2}
\end{figure}

The critical distance in large 1D arrays saturates to $\dcrit^{\text{1D}} \simeq 0.3\lambda_0$, as we demonstrate below. Jump operators in infinite 1D arrays can be written as spin waves
$\jop_{k_z} = 1/\sqrt{N}\sum_{i=1}^N \mathrm{e}^{\ii k_z z_i} \sge^i,$
where we assume the array to lie along the $z$-axis. In the above expression, $k_z$ is the wavevector and $z_i$ are the atomic positions. As $N\rightarrow\infty$, the finite set $\{k_z\}$ becomes a continuum. The decay rates can be calculated analytically for a perfect array by taking the Fourier transform of the imaginary part of the Green's function~\cite{Asenjo17PRX}, and read
\begin{subequations}
\begin{gather}
\frac{\Gamma_{\text{1D},\parallel}(k_z)}{\Gamma_0} = \frac{3\pi}{2k_0 d} \sum\limits_{g_z} \left( 1 - \frac{\left(k_z + g_z\right)^2}{k_0^2}\right),\\
\frac{\Gamma_{\text{1D},\perp}(k_z)}{\Gamma_0} = \frac{3\pi}{4k_0d} \sum\limits_{g_z} \left( 1 + \frac{\left(k_z + g_z\right)^2}{k_0^2}\right).
\end{gather}
\end{subequations}
The summations run over reciprocal lattice vectors $g_z = 2\pi n/ d, \;\forall\; n \in \mathbb{Z}$ that satisfy the condition $|g_z + k_z| \leq k_0=\omega_0/c$. These reciprocal lattice vectors correspond to scattering processes outside the first Brillouin zone. A process at $k_z + g_z$ corresponds to a discrete number of additional oscillations between sites. Therefore the processes are locally equivalent and can be folded back onto the first Brillouin zone. The sum is restricted to modes that lie inside the light cone. As shown in Fig.~\ref{Figure2}(a), for $|k_z| > k_0$, no value of $g_z$ satisfies the above condition, and modes are completely dark.

In the $N\rightarrow\infty$ limit, the condition for superradiance in 1D is recast as
\begin{align}
\int  \frac{\Gamma^2_\text{1D}(k_z)}{\Gamma_0^2} \;\mathrm{d}k_z> \frac{4\pi}{d}.
\end{align}
Integrating over the first Brillouin zone, the critical distances for both polarizations are found to be
\begin{subequations}\label{1Dcrit}
\begin{gather}
d_\text{critical}^{\text{1D}, \parallel}= \frac{3}{10} \lambda_0=0.3\lambda_0, \\
d_\text{critical}^{\text{1D}, \perp}= \frac{21}{80} \lambda_0=0.2625\lambda_0 .
\end{gather}
\end{subequations}
We demonstrate in Appendix B that the condition for superradiance is only met within the first Brillouin zone. These are therefore hard bounds on Dicke superradiance for 1D arrays.

The values of $\dcrit$ derived in the infinite limit are corroborated by numerical calculations in finite arrays, as shown in Fig.~\ref{Figure2}(b). The critical distance saturates to the analytical solution even for modest atom numbers. Parallel polarization produces higher values of $\dcrit$ because some operators within the light cone are highly subradiant, which, due to the fixed trace of $\mathbf{\Gamma}$, produces more superradiant operators and increases the variance [see Fig.~\ref{Figure2}(a)].
 
\begin{figure*}
    \centering
    \includegraphics[width=0.95\textwidth]{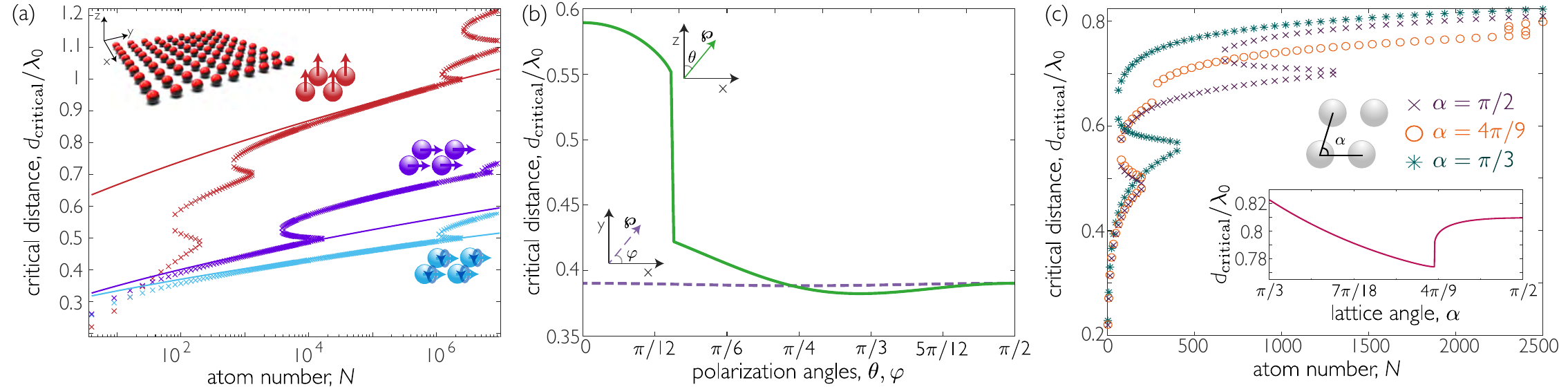}
    \caption{The critical distance increases with atom number in 2D. (a) Scaling of the critical distance with atom number for square 2D arrays for three polarizations (circular in the plane of the array, in blue; linear in the plane of the array, in purple; and out-of-plane, in red). The solid lines show guides to the eye following the analytical expression for an infinite array, i.e., Eq.~\eqref{2Dcrit} with only two free parameters. We set the parameter $C=2/15$ for the in-plane circular polarization below $d < 0.5\lambda_0$, and it remains fixed for all other lines. We set the parameter $B=4$ for perpendicular polarization and take zero for the others. The coefficients $a$ and $b$ are $b=1/8$ for circular in-plane polarization and $b=3/16$ for linear in-plane polarization, with $a=1$ for both. For out-of-plane polarization $a=0$ and $b=1/2$. (b) Maximum critical distance as a function of polarization angle, for in-plane (purple) and out-of-plane (green) polarization, for $N=10^2=100$ atoms. (c) Scaling of the critical distance with atom number for square (purple crosses), rhombic (orange circles) and triangular (teal stars) lattices. (inset) Maximum critical distance for $N=50^2=2500$ atoms in a 2D lattice as a function of lattice angle. In both plots, atoms are polarized out-of-plane. \label{Figure3}}
 \end{figure*}  

\section{2D arrays}

One-dimensional arrays are unique, as superradiance for two dimensions and above occurs for any distance in the thermodynamic limit. For 2D arrays as $N\rightarrow\infty$, the spatial profiles of the jump operators admit a description in terms of plane waves, i.e.,
$
\jop_{\kb} = 1/\sqrt{N} \sum_{i=1}^N \mathrm{e}^{i\kb\cdot\rb_i} \sge^i,
$
where $\kb$ is a two-component wavevector that lives in the plane of the array. As in 1D, decay rates can be found analytically~\cite{Asenjo17PRX}
\begin{subequations}
\begin{align}
\frac{\Gamma_{\text{2D} ,\perp}(\kb)}{\Gamma_0} &= \frac{3\pi}{k_0^3d^2} \sum\limits_{\gb} \frac{|\kb+\gb|^2}{\sqrt{k_0^2 - |\kb + \gb|^2}},\\
\frac{\Gamma_{\text{2D} ,\parallel}(\kb)}{\Gamma_0} &= \frac{3\pi}{k_0^3d^2}\sum\limits_{\gb} \frac{k_0^2 - |(\kb + \gb)\cdot \db|^2}{\sqrt{k_0^2 - |\kb+ \gb|^2}}, 
\end{align}
\end{subequations}
where the summations run over all reciprocal lattice vectors $\gb = \{ 2\pi n/d,  2\pi m/d\},\;\forall\; n,m \in \mathbb{Z}$ that satisfy $|\kb + \gb| \leq k_0$. 
\color{black} 

The condition for a superradiant burst in 2D is
\begin{equation}
\int \frac{\Gamma_\text{2D}^2(\kb)}{\Gamma_0^2}  \;\dkb> \frac{8\pi^2}{d^2}.
\end{equation}
In the first Brillouin zone, the integral of the square of the decay rates for out-of-plane polarization is
\begin{equation}
\int \left( \frac{\Gamma_{\text{2D},\perp}(\kb)}{\Gamma_0} \right)^2 \dkb = \frac{9\pi^2}{k_0^6d^4} \int\limits_0^{2\pi}\mathrm{d}\theta \int\limits_0^{k_0} \frac{k^5}{k_0^2 - k^2} \mathrm{d}k,
\end{equation}
which diverges logarithmically as $k\rightarrow k_0$. The same is true for the integral for in-plane polarization. While we only perform the integral in the first Brillouin zone here, an identical divergence occurs for $|\kb + \gb|\rightarrow k_0$ in whichever Brillouin zones that condition is met. We isolate the divergence by integrating up to $k_0(1-\varepsilon)$, where $\varepsilon\rightarrow 0$ is a small deviation from $k=k_0$. A large finite array of atom number $N$ samples each dimension of the first Brillouin zone with frequency $N^{1/n}$, with $n$ being the array dimensionality. By taking $\varepsilon=C\lambda_0/d\sqrt{N}$ (with $C$ being a constant), we show analytically in Appendix B that for large $N$ the critical distance in 2D scales as
\begin{equation}
d_\text{critical}^{\text{2D}}\simeq \lambda_0\sqrt{\alpha +\beta\ln{N}},\label{2Dcrit}
\end{equation}
where $\alpha=9(a-3b-2b\ln{2C}+B)/32\pi$ and $\beta=9b/32\pi$ are reduced constants that depend on the polarization via $a$ and $b$, the sampling constant $C$ that depends only on the lattice geometry\color{black}, and a factor $B$ that depends on the set of relevant reciprocal lattice vectors [see Appendix~B for the detailed expression]. This result agrees with numerical findings, as shown in Fig.~\ref{Figure3}(a).

We have demonstrated that in 2D the critical distance scales sublogarithmically with atom number. This, however, should not be understood as having superradiance for any distance for infinite arrays, as multiple approximations our model relies on break down in this limit. For instance, the derivation of the master equation [Eq.~\eqref{masterequation}] is done under the Born-Markov approximation, which assumes that the maximum interatomic distance is small enough that one can ignore the propagation time of photons between atoms. As $Nd\rightarrow\infty$, this approximation fails. Nevertheless, the critical distance for any large number of atoms will be significantly higher than in 1D.

While numerical calculations confirm the sub-logarithmic scaling of $\dcrit$, Fig.~\ref{Figure3}(a) shows that the maximum critical distance does not increase smoothly. This is due to the geometric resonances discussed above, which produce revivals in the variance and disconnected regions of superradiance~\cite{Masson22}. Atoms with in-plane polarization produce smaller values of $\dcrit$ because for the linear case far-field emission is forbidden in one direction in the plane, and in the circular case it is reduced in both. The $1/r$ terms are less significant, diminishing revivals in the variance and slowing the growth of maximum $\dcrit$ with $N$. This is evident in Fig.~\ref{Figure3}(a), where the discontinuity for the in-plane polarized case is much smaller in size and occurs at much higher $N$ than for the out-of-plane polarized case.

As the polarization vector is rotated towards the plane, there is a sudden drop in the maximum $\dcrit$, as the revivals in the variance diminish until they can no longer sustain a superradiant burst, as shown in Fig.~\ref{Figure3}(b). Rotating the polarization in plane yields little change, as the choice of axis is unimportant for large arrays.

As shown in Fig.~\ref{Figure3}(c), the geometry of the lattice has only a minor impact on the critical distance. For large $N$, a triangular lattice maximizes $\dcrit$, as this geometry minimizes all periodic distances between neighboring atoms. However, the scaling appears to be the same in all cases. The number of discontinuities in the maximum $\dcrit$ changes with geometry. For a square lattice, there are two discontinuities in the region $d \in [0.5\lambda_0,\lambda_0]$, one associated to the nearest neighbor distance, and one for the diagonal displacement across the unit cell. For a rhombic lattice, there are three discontinuities, as the diagonal displacements are different. For a triangular lattice, there is only one, as all distances of the unit cell are the same.

\color{black}
    
\begin{figure}[b!]
    \centering
    \includegraphics[width=0.475\textwidth]{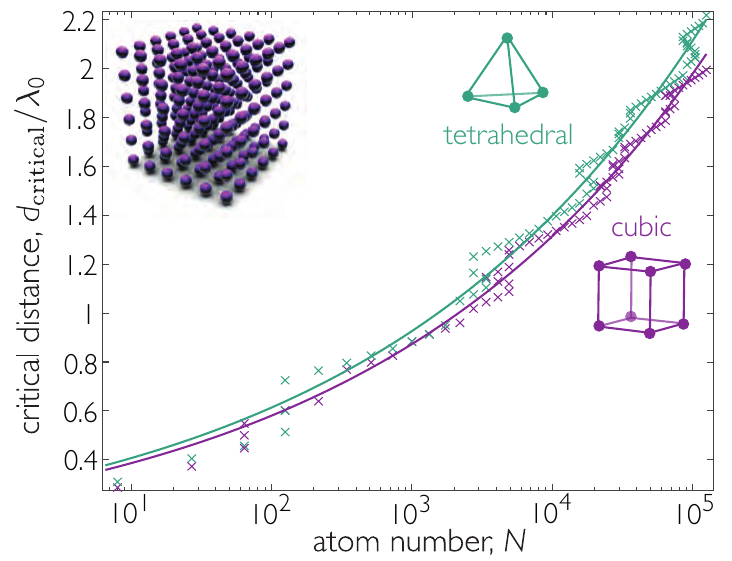}
    \caption{The maximum critical distances diverges as a power law in 3D, as shown for atoms arranged in cubic (purple) and tetrahedral (turquoise) lattices. In both cases, the transition polarization is parallel to one of the main axes of the array. The line represents the best fits to the function $d_\text{critical}/\lambda_0= q N^p$, with $q=\{0.255, 0.268\}$ and $p=\{0.178, 0.180\}$ for the cubic and the tetrahedral geometries, respectively.}
    \label{Figure4}
\end{figure}  

\section{3D arrays}

Infinite 3D arrays are fundamentally different, in that they cannot radiate to far-field. However, a regularization factor $i\Delta k_0^2$, with $\Delta \rightarrow 0^+$, is required to avoid a divergence in the Green's function at $|\kb + \gb| = k_0$~\cite{Antezza09,Brechtelsbauer21}. This leads to decay rates of the form
\begin{equation}
    \frac{\Gamma_{\text{3D}}(\kb)}{\Gamma_0} = \frac{6\pi}{k_0d^3} \sum\limits_{\gb} \frac{\Delta(k_0^2-|(\kb + \gb)\cdot\db|^2)}{\left(k_0^2 - |\kb + \gb|^2\right)^2 +\Delta^2k_0^4},
\end{equation}
where the summation runs over all reciprocal lattice vectors $\gb = \{ 2\pi n/d,  2\pi m/d, 2\pi l/d\},\;\forall\; n,m,l \in \mathbb{Z}$ that satisfy $|\kb + \gb| \leq k_0$.
\color{black}

We analytically find that $\dcrit$ diverges as $N^{1/6}$ for 3D arrays (see Appendix B). This agrees with the numerical findings, as shown in Fig.~\ref{Figure4}. As in 2D, the optimal geometry appears to be the one that optimizes packing efficiency and thus minimizes all displacement vectors. Here, this is the tetrahedral lattice, for which $\dcrit \approx 2.3 \lambda_0$ for a $N=50^3$ array. As in 2D, there does not seem to be any saturation of $\dcrit$ as $N\rightarrow\infty$, indicating a divergent critical distance in 3D.

\section{Conclusions}

\begin{figure}
    \centering
    \includegraphics[width=0.475\textwidth]{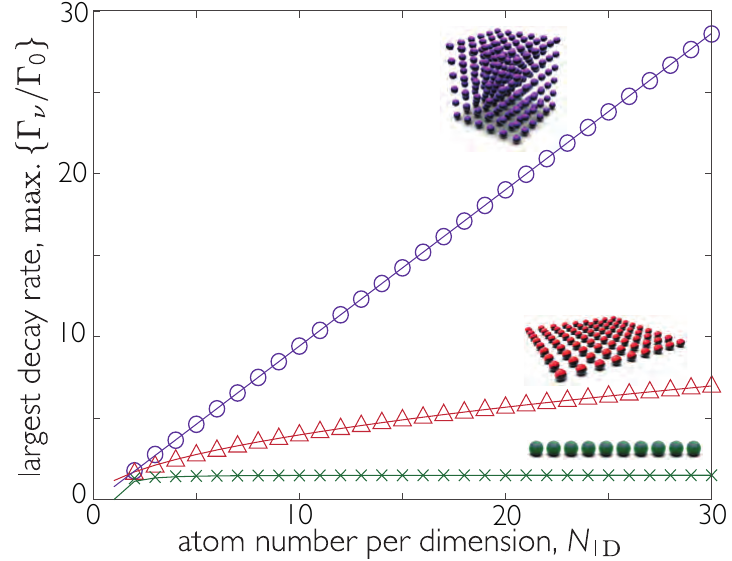}
    \caption{The largest decay rate saturates in 1D, and grows with $N$ in 2D and 3D. Scaling of the largest decay rate in different dimensionalities with $d= 0.5\lambda_0$. In all cases, atoms are polarized along one of the axis of the array and the lines of best fit are based on finite samplings of the infinite expressions. The scalings are $\sim(1 - c/N_{\text{1D}}^2)$ in 1D, $\sim \sqrt{N_{\text{1D}}}$ in 2D, and $\sim N_{\text{1D}}$ in 3D, with $N_{\mathrm{1D}} =N^{1/n}$ being the number of atoms along each dimension and $c$ being a constant.}
    \label{Figure5}
\end{figure}

Array dimensionality determines the physical origin of Dicke superradiance. In 1D, superradiance occurs due to destructive interference, which produces dark decay channels with a very suppressed radiative decay. The bright channels have decay rates that are not large and do not increase with atom number [as shown in Fig.~\ref{Figure5}], as there are not enough atoms in a chain to produce robust constructive interference. In 2D and above, superradiance occurs because of constructive interference, which results in large decay rates for channels with wave-vectors $|\textbf{k}|\simeq k_0$ [which grow with atom number, as shown in Fig.~\ref{Figure5}]. Destructive interference is not behind Dicke superradiance in high-dimensions because there are no dark decay rates beyond a certain distance, in contrast to superradiance, which occurs for any distance if the array is large enough. While Dicke superradiance is a transient phenomenon, there are similarities with phase transitions in that long-range order emerges for high dimensionalities, and 2D is a ``lower critical dimension'' (1D should not display Dicke superradiance, but there is ``residual'' superradiance because of destructive interference).

Finally, we note that our method is limited to identifying the minimal conditions for the presence of a burst, but not the properties of the burst. Information about, for example, the scaling of the peak emission requires a different approach~\cite{Robicheaux21PRA_HigherOrderMeanField,Rubies21arxiv}. We also consider only the total photon emission (i.e., integrating the radiation over all directions). As we discuss in Ref.~\cite{Masson22}, the critical distance may be higher if the emitted light is measured only in particular directions. Our algebraic technique has been recently adapted for that purpose~\cite{Robicheaux21PRA_Directional}.

In conclusion, we have shown that superradiant bursts can exist well beyond the regime first considered by Dicke. By calculating the maximum interatomic distance at which a superradiant burst occurs, we find that superradiance persists to much larger distances in higher-dimensional arrays, as it occurs due to constructive interference processes. Current state-of-the-art optical tweezer array and optical lattice experiments already operate well below these bounds~\cite{Rui20,Glicenstein20}, and superradiance could thus be observed in such systems. Arrays of solid-state emitters hosted in 2D materials~\cite{Palacios17,Proscia18,Li21} or in bulk crystals~\cite{Kornher16,Sipahigil16} can also be employed to explore this physics. Our theoretical methods can be applied to studies of collective emission with atoms with more complex level structure~\cite{Lin12,Sutherland17,PineiroOrioli22}, and atoms coupled to other reservoirs, such as nanophotonic structures~\cite{Goban15,Solano17}.

\textbf{Acknowledgments} -- We are thankful for insightful discussions with S. Yelin and O. Rubies Bigorda. The work on 2D arrays has been supported by Programmable Quantum Materials, an Energy Frontier Research Center funded by the U.S. Department of Energy (DOE), Office of Science, Basic Energy Sciences (BES), under award No. DE-SC0019443. The work for other dimensionalities has been supported by the Alfred P. Sloan Foundation through a Sloan Research Fellowship, the David and Lucile Packard Foundation, and the National Science Foundation CAREER Award (No. 2047380).

\textit{Note added} -- During the writing phase of this manuscript, we became aware of related work by F. Robicheaux~\cite{Robicheaux21PRA_Directional}.

\appendix

\section{Calculating the variance of the decay rates in $O(N)$ steps}\label{appa}

Here we present the algorithm that allows us to find the variance of the decay rates in $O(N)$ steps, instead of the $O(N^3)$ scaling that would result from employing common algorithms to find the eigenvalues of an $N\times N$ matrix. Let $\mathbf{A}$ be the normalized matrix $\mathbf{\Gamma}/\Gamma_0$ with components $A_{ij}$ and eigenvalues $\{\lambda_i\}$. The matrix $\Ab$ inherits symmetry properties from the Green's tensor. Due to reciprocity, $G_{\alpha\beta}(\rb_i,\rb_j,\omega_0)=G_{\beta\alpha}(\rb_j,\rb_i,\omega_0)$ and since all atoms share a common quantization axis, $\alpha=\beta$ for all $i,j$. Therefore, $\Ab$ is real and symmetric. The diagonal elements of $\Gab$ correspond to the single atom decay rate ($\Gamma^{ii}=\Gamma_0$) such that
\begin{equation}
    \Tr \Ab =\sum_{i=1}^N\lambda_i = N.
\end{equation}
Let $\Db$ be the diagonalization of $\Ab$ such that $\Ab=\Pb \Db \Pb^{-1}$ (the spectral theorem ensures $\Ab$ can be diagonalized as it is real and symmetric), with $\Db$ being a diagonal matrix with diagonal elements $\lambda_1,\dots,\lambda_N$. Therefore,
\begin{equation}
    \Db^2 = 
    \begin{pmatrix}
    \lambda_1^2 & & \\
     & \ddots & \\
     & & \lambda_N^2\\
    \end{pmatrix},
\end{equation}
with the trace of $\Db^2$ being all we need to obtain the variance. Since
\begin{equation}
    \Ab^2=\Pb\Db\Pb^{-1}\Pb\Db\Pb^{-1}=\Pb\Db^2\Pb^{-1},
\end{equation}
we only need to calculate the trace of $\Ab^2$, since $\Tr \Ab^2 = \Tr \Db^2$. Multiplying matrices is computationally costly, but we can avoid that operation by realizing that
\begin{equation}
    \Tr \Ab^2 = \sum_{i=1}^N \left(\sum_{j=1}^N A_{ij}A_{ji}\right)=\sum_{i,j=1}^N A_{ij}^2,
\end{equation}
as $\Ab$ is real and symmetric. The sum of squared decay rates is then
\begin{align}
    &\sum_{\nu=1}^N \left(\frac{\Gamma_\nu}{\Gamma_0}\right)^2 = \sum_{i,j=1}^N \left(\frac{\Gamma^{ij}}{\Gamma_0}\right)^2= N+  2 \sum_{i=1,j>i}^N \left(\frac{\Gamma^{ij}}{\Gamma_0}\right)\notag\\
    &= N +2\sum_\beta n_\beta \left(\frac{\Gamma_\beta}{\Gamma_0}\right)^2,
     \label{ONexp}
\end{align}
where we sum over all different displacement vectors $\{\mathbf{s}_\beta\}$, $\Gamma_\beta$ is the dissipative interaction rate between atoms separated by displacement vector $\mathbf{s}_\beta$, and $n_\beta$ is the number of repetitions of $\mathbf{s}_\beta$ in the array. For arbitrary geometries (where all vectors are unique), this scales as $O(N^2)$. However, in ordered arrays, the multiplicity of most displacement vectors is larger than one (for example, in a 1D array, $\textbf{s}_2\equiv\rb_{1,3}=\rb_{2,4} = - \rb_{5,3}=...$). This reduces the complexity of calculating the sum of squared decay rates to the number of different displacement vectors.

In ordered arrays, the sum in $\beta$ has $O(N)$ terms and the variance is calculated in $O(N)$ steps. The atoms occupy a grid such that all displacement vectors can be expressed as integer numbers of discrete steps in each dimension. To this end, we define a dimensionless vector $\mathbf{d}_\beta = \{a_\beta,b_\beta,c_\beta\}$ with $a_\beta, b_\beta, c_\beta$ such that $\mathbf{s}_\beta = a_\beta \hat{\mathbf{s}}_1 + b_\beta \hat{\mathbf{s}}_2 + c_\beta \hat{\mathbf{s}}_3$ with $\hat{\mathbf{s}}_{1,2,3}$ the three vectors that describe the unit cell. We define $N_\text{1D}$ as the number of atoms along a one-dimensional slice in the array: in 1D $N_\text{1D} = N$ and $b_\beta=c_\beta=0$, in 2D $N_\text{1D}=\sqrt{N}$ and $c_\beta=0$, and in 3D $N_\text{1D}=N^{1/3}$.

\textit{Counting in 1D}-- All displacement vectors can be identified by considering one end atom. The shortest displacement vector is that of nearest neighbors: $|\mathbf{d}_\beta| = 1$. This appears $N-1$ times, as $\rb_{1,2} = \rb_{2,3}= \dots =\rb_{N-1,N}$. The largest displacement vector is that between the two end atoms, $|\mathbf{d}_\beta| = N-1$, which appears only once. More generally, a displacement vector of $a_\beta$ is repeated $n_\beta = N_{\text{1D}} - a_\beta$ times.

\begin{figure}
    \centering
    \includegraphics[width=0.375\textwidth]{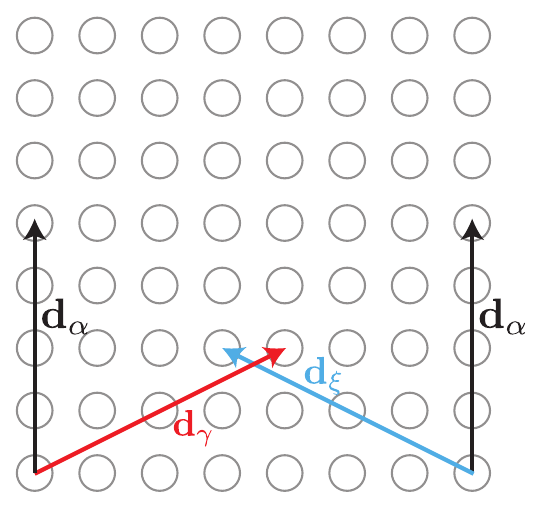}
    \caption{Schematic of different displacement vectors in a square lattice.}
    \label{Figure6}
\end{figure}

\textit{Counting in 2D}-- All displacement vectors can be identified by considering two adjacent corners. The displacement vectors between the first corner and all other atoms provides all displacement vectors of the form $\mathbf{d}_\beta = \{a_\beta,b_\beta\}$ with $0 \leq a_\beta,b_\beta \leq N_{\text{1D}}-1$ (excluding $\{0,0\}$, which are counted in the factor of $N$ in Eq.~\eqref{ONexp}). However, this does not include vectors where $a_\beta$ (or equivalently $b_\beta$) are negative. For example, in Fig.~\ref{Figure6}, displacements from the bottom left corner yield $\mathbf{d}_\alpha$ and $\mathbf{d}_\gamma$, but they do not include $\mathbf{d}_\xi$. To complete the set of possible displacement vectors we require those given by a second-corner: $\mathbf{d}_\beta = \{a_\beta,-b_\beta\}$ (or equivalently $\mathbf{d}_\beta = \{-a_\beta,b_\beta\}$, depending on the choice of corner) with $1 \leq a_\beta, b_\beta \leq N_{\text{1D}}-1$. Note that here $a_\beta, b_\beta \neq 0$, since those vectors are already included in those defined by the first corner. To count repetitions, we note that, as in 1D, a vector which covers $a_\beta$ sites in the first dimension fits $N_{\text{1D}} - a_\beta$ times. The same holds for the second dimension so the total repetitions is
\begin{equation}
n_\beta = \left( N_\text{1D} - a_\beta \right) \left( N_\text{1D} - b_\beta \right).
\end{equation}

\textit{Counting in 3D}-- All displacement vectors can be identified by considering the four corners of one face. As in 2D, the first corner provides all displacement vectors of the form $\mathbf{d}_\beta = \{a_\beta,b_\beta,c_\beta\}$ with $0 \leq a_\beta, b_\beta, c_\beta \leq N_{\text{1D}}-1$ (again excluding $\{0,0,0\})$. The other three corners provide the rest of the set, with care being taken to not repeat vectors. The number of repetitions is then
\begin{equation}
n_\beta = \left( N_\text{1D} - a_\beta \right) \left( N_\text{1D} - b_\beta \right)\left( N_\text{1D} - c_\beta\right).
\end{equation}

\section{Calculations for infinite arrays\label{inf}}

The analytical expressions for the decay rates for infinite 1D, 2D and 3D arrays are required to understand the behavior of large atomic arrays. The spectral representation of the dyadic Green's  function is found as the solution of the Fourier transform of the electromagnetic wave equation \cite{chew1995waves}
\begin{equation}
    \tilde{G}_0(\kb,\rb')=\frac{k_0^2\mathds{1}-\kb\otimes\kb}{k_0^2\,(|\kb|^2-k_0^2)}e^{-\ii\kb\cdot\rb'},
\end{equation}
from which one readily finds
\begin{equation}
    G_0(\rb,\rb')=\frac{1}{(2\pi)^3}\int\limits_{-\infty}^{\;+\infty}\frac{k_0^2\mathds{1}-\kb\otimes\kb}{k_0^2\,(|\kb|^2-k_0^2)}e^{-\ii\kb\cdot(\rb-\rb')}\mathrm{d}\kb.\label{eq:Gten}
\end{equation}
For 1D and 2D, we choose $k_x$ as a preferred direction perpendicular to the array. By virtue of Jordan's lemma and Cauchy's theorem, we can integrate the $k_x$ dependence accounting for the residue contributions at the poles $\pm k_{0x}$, where $k_{0x} = \sqrt{k_0^2-k_y^2-k_z^2}$ and find
\begin{equation}
    G_0(\rb,\rb')=\frac{\ii}{8\pi^2k_0^2}\int\limits_{-\infty}^{\;+\infty}\frac{k_0^2\mathds{1}-\bar{\kb}\otimes\bar{\kb}}{k_{0x}}e^{\ii\kb_s(\rb_s-\rb_s')+\ii k_{0x}\abs{x}}\mathrm{d}\kb_s,
\end{equation}
where $\kb_s=k_y\hat{y}+k_z\hat{z}$, $\rb_s=y\hat{y}+z\hat{z}$, and $\bar{\kb}=\kb_s + \text{sign}(x)k_{0x}\hat{x}$. We set both the 1D chain and the 2D array at the plane $x=0$ and evaluate the Green's function at atomic positions (with $x=0$). Making use of the Dirac delta representation in $n$ dimensions,
\begin{equation}
    \sum_{\rb_i \in \text{lattice}} e^{\ii\kb\cdot\rb_i}=\left(\frac{2\pi}{d}\right)^n\sum\limits_{\substack{\gb \in \text{reciprocal}\\ \text{lattice}}}\delta^{(D)}(\kb-\gb),
\end{equation}
it is possible to express $\tilde{G}_0(\kb)$ as a sum over reciprocal lattice vectors $\gb$. For a 1D chain along the $\hat{z}$ axis, we obtain 
\begin{equation}
    \tilde{G}_0(\kb)=\frac{\ii}{8\pi^2}\frac{2\pi}{k_0^2d}\sum_{g_z} \int\limits_{-\infty}^{\infty}\frac{k_0^2\mathds{1}-\bar{\kb}_1\otimes\bar{\kb}_1}{k_{1x}}\mathrm{d}k_y,
\end{equation}
where $k_{1x}=\sqrt{k_0^2-k_y^2-(k_z+g_z)^2}$ and $\bar{\kb}_1=\text{sign}(x)k_{1x}\hat{x} + k_y\hat{y} + (k_z + g_z)\hat{z}$. In the case of a 2D square lattice
\begin{equation}
    \tilde{G}_0(\kb)=\frac{\ii}{8\pi^2}\left(\frac{2\pi}{k_0d}\right)^2\sum_{\gb} \frac{k_0^2\mathds{1}-\bar{\kb}_2\otimes\bar{\kb}_2}{k_{2x}},
\end{equation}
where $k_{2x}=\sqrt{k_0^2-(k_y+g_y)^2-(k_z+g_z)^2}$ and $\bar{\kb}_2=\text{sign}(x)k_{2x}\hat{x} + (k_y+g_y)\hat{y} + (k_z + g_z)\hat{z}$. For the 3D lattice we can directly use the Dirac delta representation from Eq.~(\ref{eq:Gten}) to find
\begin{equation}
    \tilde{G}_0(\kb)=\frac{1}{k_0^2d^3}\sum_{\gb}\frac{k_0^2\mathds{1}-(\kb+\gb)\otimes(\kb+\gb)}{|\kb+\gb|^2-k_0^2+\ii\Delta k_0^2},
\end{equation}
where we have introduced the regularization factor $\ii\Delta k_0^2$ to avoid the divergence at $|\kb+\gb|=k_0$, with $\Delta\to 0^+$. Then, the collective decay rates can be evaluated, as
\begin{equation}
    \frac{\Gamma_\kb}{\Gamma_0}=\frac{6\pi}{k_0}\db\cdot\Im\tilde{G}_0(\kb)\cdot\db,
\end{equation}
leading to the expressions provided in the main text.

As expected, for all dimensionalities
\begin{equation}
    \int\limits_{BZ1} \frac{\Gamma(\kb)}{\Gamma_0}\;\mathrm{d}\kb =\left(\frac{2\pi}{d}\right)^n,
\end{equation}
where $n$ is the dimension of the array. 

\subsection{1D arrays with $d < 0.5\lambda_0$\label{infBZ1}}

Here, we consider a 1D array with $d/\lambda_0<0.5$, where all non-zero decay rates are contained within the first Brillouin zone and the only non-zero term in the sum is $g_z=0$. Performing the integrals in k-space for both polarizations, we find
\begin{subequations} 
\begin{gather} 
\int\limits_{BZ1} \left(\frac{\Gamma_{\text{1D} ,\perp}(k_z)}{\Gamma_0}\right)^2  \;\dk
= \frac{21}{40k_0}\left(\frac{2\pi}{d}\right)^2,\label{eq:perpsquare}\\
\int\limits_{BZ1}  \left(\frac{\Gamma_{\text{1D} , \parallel}(k_z)}{\Gamma_0}\right)^2 \;\dk
= \frac{3}{5k_0}\left(\frac{2\pi}{d}\right)^2.\label{eq:parsquare}
\end{gather}
\end{subequations} 
The condition for a minimum burst yields the critical distances of Eq.~\eqref{1Dcrit} in the main text.

\subsection{1D arrays with $d > 0.5\lambda_0$\label{infBZn}}

Here we demonstrate that the critical distances that we have found in the previous section are the only possible ones. For $0.5<d/\lambda_0<1$, there are three possible values of $g_z=\{-2\pi/d, 0, 2\pi/d\}$ that contribute to the decay rates. The integrals of the square of the decay rates for both polarizations yield

\begin{widetext}
\begin{subequations}
\begin{align}
\int\limits_{BZ1}&\left(\frac{\Gamma_{\text{1D} ,\perp}(k_z)}{\Gamma_0}\right)^2\,\dk = \frac{9\pi^2}{16k_0^2d^2}\int\limits_{-\pi/d}^{\pi/d} \sum_{g_z}\left(1+\frac{\left(k_z+g_z\right)^2}{k_0^2}\right)^2+\sum_{\substack{g_z,g_z'\\ g_z\neq g_z'}}\left(1+\frac{\left(k_z+g_z\right)^2}{k_0^2}\right)\left(1+\frac{\left(k_z+g_z'\right)^2}{k_0^2}\right)\,\dk\nonumber\\
&=\frac{2\pi^2}{k_0d^2}\left(-\frac{3\lambda_0^5}{5d^5}-\frac{3\lambda_0^3}{d^3}+\frac{6\lambda_0^2}{d^2}-\frac{9\lambda_0}{2d} + \frac{63}{20}\right),\\
\int\limits_{BZ1}&\left(\frac{\Gamma_{\text{1D} ,\parallel}(k_z)}{\Gamma_0}\right)^2\,\dk =\frac{2\pi^2}{k_0d^2}\left(-\frac{12\lambda_0^5}{5d^5}+\frac{12\lambda_0^3}{d^3}-\frac{12\lambda_0^2}{d^2}-\frac{18}{5}\right).
\end{align}
\end{subequations}
\end{widetext}
These expressions do not satisfy the minimal condition for a superradiant burst [i.e., Eq.~\eqref{condition}] anywhere in the region $0.5 < d/\lambda_0 < 1$.

For larger interatomic separations, we can upper bound the integral to prove that the variance is smaller than unity always. For perpendicular polarization we bound the double sum by
\begin{widetext}
\begin{align}\label{eq:bound1D}
&\sum_{g_z, g_z'}\Biggl(1+\frac{\left(k_z+g_z\right)^2}{k_0^2}\Biggr)\left(1+\frac{\left(k_z+g_z'\right)^2}{k_0^2}\right) < \sum_{g_z}\left(1+\frac{\left(k_z+g_z\right)^2}{k_0^2}\right)^2 + 2N_{g_z}\left(1+\frac{\left(k_z+g_z\right)^2}{k_0^2}\right)^2,
\end{align}
\end{widetext}
where $N_{g_z}$ is a natural number such that $g_z=\pm N_{g_z} 2\pi/d$. The inequality is produced by upper bounding each of the products in the sum as the square of whichever term is larger. This allows us to bound the integral as
\begin{align}
\int\limits_{BZ1}\left(\frac{\Gamma_{\text{1D} ,\perp}(k_z)}{\Gamma_0}\right)^2\,\dk <\frac{21\pi^2}{5k_0d^2}+\frac{63\pi}{48d},
\end{align}
which satisfies the condition for a superradiant burst for $d/\lambda_0<0.7814$. Since we have already shown that the superradiant burst cannot be exhibited for 1D arrays with $0.2625 < d/\lambda_0 < 1$, this means that $d < 0.2625\lambda_0$ is the only region in which a superradiant burst can be produced.

Similarly, for parallel polarization we bound the integral by
\begin{align}
\int\limits_{BZ1}\left(\frac{\Gamma_{\text{1D} ,\parallel}(k_z)}{\Gamma_0}\right)^2\,\dk <\frac{24\pi^2}{5k_0d^2}+\frac{3\pi}{4d},
\end{align}
which satisfies the condition for a superradiant burst if $d/\lambda_0<0.8571$. Again, we have shown previously that a superradiant burst cannot be produced for $0.3 < d/\lambda_0 < 1$ such that $d<0.3\lambda_0$ is the only region a superradiant burst can be produced.

\subsection{2D arrays with $d<0.5\lambda_0$\label{inf2D}}\label{ap2d}

For 2D arrays with $d<0.5\lambda_0$, the non-zero decay rates entirely lie within the first Brillouin zone and the only possible terms are those where $\gb = \{0,0\}$. Performing the integral yields
\begin{equation}
    \int\limits_{BZ1}  \left(\frac{\Gamma_{\text{2D}}(\kb)}{\Gamma_0}\right)^2 \; \dkb =\frac{9\pi\lambda_0^2}{4 d^4}\left(a+\frac{4b}{k_0^4}\int\limits_0^{k_0} \frac{k^5}{k_0^2-k^2}\mathrm{d}k\right),
\end{equation}
where $a$ and $b$ depend on the polarization. For perpendicular polarization $a=0$ and $b=1/2$, for in-plane linear polarization $a=1$ and $b=3/16$, and for in-plane circular polarization $a=1$ and $b=1/8$. The integral is performed over all non-zero decay rates, which form a disc, $D_{k_0}$, centred on zero with radius $k_0$. The integral diverges as $|\kb|\rightarrow k_0$, but we characterize the scaling of the divergence with atom number. We isolate the divergence by integrating up to $k_0(1-\varepsilon)$, where $\varepsilon \rightarrow 0$ represents a small deviation from $k=k_0$:
\begin{widetext}
\begin{align}
    &\frac{1}{k_0^4}\int\limits_0^{k_0}\frac{k^5}{k_0^2-k^2}\mathrm{d}k = \lim_{\varepsilon\to 0} \int\limits_0^{1-\varepsilon}\frac{x^5}{1-x^2}\mathrm{d}x = \lim_{\varepsilon\to 0} -\frac{\varepsilon^4}{4} +\varepsilon^3 -2\varepsilon^ 2 +\varepsilon -\frac{3}{4} -\frac{1}{2}\ln{(2\varepsilon-\varepsilon^2)}.
\end{align}
\end{widetext}
A finite array approximately samples each dimension of the Brillouin zone with spacing $1/\sqrt{N}$, so  $\varepsilon=C\lambda_0/d\sqrt{N}$, with $C$ being a constant. By considering the condition given in Eq.~\eqref{condition}, we find that $\dcrit$ satisfies the transcendental equation
\begin{widetext}
\begin{equation}\label{trans}
    \frac{9\lambda_0^ 2}{16\pi d^2}\left[a+4b\left(-\frac{C^4\lambda_0^4}{4d^4N^2} +\frac{C^3\lambda_0^3}{d^3N^{3/2}} -\frac{2C^2\lambda_0^2}{d^2N} +C\frac{\lambda_0}{d\sqrt{N}} -\frac{3}{4} +\frac{1}{2}\ln{\left(\frac{1}{C}\frac{d^2N/\lambda_0^2}{2d\sqrt{N}/\lambda_0-C}\right)}\right)\right]-2=0.
\end{equation}
\end{widetext}
In the limit $N\rightarrow\infty$, $1/\sqrt{N}d \rightarrow 0$, the condition can be approximated as
\begin{equation}
    d_\text{critical}^{\text{2D}}\simeq \frac{3\lambda_0}{4\sqrt{2\pi}}\sqrt{a-3b-2b\ln{2C}+b\ln{N}}
\end{equation}
where we also assume $\ln d / \lambda_0 \ll \ln N$.

\subsection{2D arrays with $d> 0.5\lambda_0$}

For $d>0.5\lambda_0$, there are several values of $\gb$ that can satisfy $|\kb + \gb| \leq k_0$, and the integral of the squared decay rates becomes
\begin{widetext}
\begin{align}
    \left(\frac{\Gamma_{\text{2D} ,\perp}(\kb)}{\Gamma_0}\right)^2 &= \frac{9\pi^2}{k_0^6d^4}\left( \sum\limits_{\gb} \frac{|\kb+\gb|^2}{\sqrt{k_0^2 - |\kb + \gb|^2}}\right)\left( \sum\limits_{\gb'} \frac{|\kb+\gb'|^2}{\sqrt{k_0^2 - |\kb + \gb'|^2}}\right)\nonumber\\
    &= \frac{9\pi^2}{k_0^6d^4}\left( \sum\limits_{\gb=\gb'}\frac{|\kb+\gb|^2}{\sqrt{k_0^2 - |\kb + \gb|^2}}\frac{|\kb+\gb'|^2}{\sqrt{k_0^2 - |\kb + \gb'|^2}} + \sum\limits_{\gb',\gb\neq \gb'} \frac{|\kb+\gb|^2}{\sqrt{k_0^2 - |\kb + \gb|^2}}\frac{|\kb+\gb'|^2}{\sqrt{k_0^2 - |\kb + \gb'|^2}}\right).
\end{align}
\end{widetext}

The first of these terms is exactly equivalent to that with $\gb=\gb'=\{0,0\}$, and so is integrated in the same manner as above. The second term has a convergent integral as both terms cannot generally diverge when $\gb \neq \gb'$. The only exception to this is a finite set of terms that have zero measure when integrated over the first Brillouin zone. When integrated, the first term will give the same result as for $d<0.5\lambda_0$ and the second term will be a function of interatomic distance, $B(d)$. However, $B(d)$ varies slowly with $d$ for each possible set of relevant reciprocal lattice vectors, and so can be considered a constant in the intervals between additional values of $\gb$ becoming relevant. The same process can be done for all polarizations, yielding the final result
\begin{equation}
    d_\text{critical}^{\text{2D}}\simeq \frac{3\lambda_0}{4\sqrt{2\pi}}\sqrt{a-3b-2b\ln{2C}+ B +b\ln{N}}.
\end{equation}

\subsection{3D arrays with $d< 0.5\lambda_0$}

For 3D arrays with $d<0.5\lambda_0$, we have
\begin{widetext}
\begin{align}
    \int\limits_{BZ1}  \left(\frac{\Gamma_{\text{3D}}(\kb)}{\Gamma_0}\right)^2 \;\mathrm{d} \mathbf{k} &=\frac{36\pi^2}{k_0^2d^6}\int\limits_0^{\infty}\int\limits_{0}^{\pi}\int\limits_0^{2\pi} \frac{\Delta^2\left(k_0^2-k^2\cos^2{\theta}\right)^2}{\left(\left(k_0^2-k^2\right)^2+\Delta^2k_0^4\right)^2}k^2\sin{\theta}\;  \mathrm{d}\phi\mathrm{d}\theta\mathrm{d}k\nonumber\\
    &=\frac{36\pi^2}{k_0^2d^6}\int\limits_0^{\infty}\Delta^2\frac{4\pi k_0^4k^2-8\pi k_0^2k^4/3+4\pi k^6/5}{\left(\left(k_0^2-k^2\right)^2+\Delta^2k_0^4\right)^2}\;\mathrm{d}k = \left(\frac{2\pi}{d}\right)^3\frac{18}{k_0^3d^3}\int\limits_0^{\infty}\Delta^2\frac{x^2-2x^4/3+x^6/5}{\left(\left(1-x^2\right)^2+\Delta^2\right)^2}\;\mathrm{d}x\nonumber\\
    &=\left(\frac{2\pi}{d}\right)^3\frac{18}{k_0^3d^3}\left(H_\Delta(+\infty)-H_\Delta(1^+) +H_\Delta(1^-) -H_\Delta(0)\right).
\end{align}
In this case, the integral diverges at $x=1$ in the limit $\Delta\to 0$. The function $H_\Delta(x)$ is defined as
\begin{align}
    H_\Delta(x)&=-\frac{1}{120}\frac{(6\Delta^2-16)x^3-(14\Delta^2-16)x}{\Delta^2+(x^2-1)^2} + \frac{1}{120}\sqrt{\frac{\ii}{\Delta-\ii}}\left(9\Delta^2-2\ii\Delta+8-\ii\frac{16}{\Delta}\right)\arctan{\left(x\sqrt{\frac{\ii}{\Delta-\ii}}\right)}+\mathrm{c.c.}
\end{align}
such that $H_\Delta(0)=0$ and
\begin{equation}
    H_\Delta(+\infty)=-\frac{1}{120}\left(\ii\frac{8\pi}{\Delta}+\ii\Delta\pi-4\pi-\frac{9}{2}\pi\Delta^2\right)\left(\sqrt{\frac{\ii}{\Delta-\ii}}-\sqrt{\frac{-\ii}{\Delta+\ii}}\right),
\end{equation}
\end{widetext}
which has a null real part and a divergent imaginary part when $\Delta\to 0$. For $H_\Delta(1^-)-H_\Delta(1^+)$, the second term is zero and the first term is divergent for $\Delta\to 0$. Therefore, the divergence of the integral arises from the first polynomial part of $H_{\Delta}(x)$. Setting $1^+=1+\varepsilon$ and $1^-=1-\varepsilon$, the real part of the integral is
\begin{widetext}
\begin{align}
    \int\limits_{BZ1}  \left(\frac{\Gamma_{\kb}}{\Gamma_0}\right)^2 \;\mathrm{d}\kb = \left(\frac{2\pi}{d}\right)^3\frac{3}{20k_0^3d^3} \left(\frac{1}{\varepsilon} +\frac{1}{2(\varepsilon+2)} +\frac{1}{2(\varepsilon-2)}\right) \simeq \left(\frac{2\pi}{d}\right)^3\frac{3}{20k_0^3d^3}\frac{1}{\varepsilon}.
\end{align}
\end{widetext}

The sampling of the reciprocal space by a finite array is proportional to the cubic root of $N$. Therefore, $\varepsilon=C\lambda_0/d N^{1/3}$. We thus find 
\begin{equation}
    \dcrit \simeq \lambda_0\sqrt{\frac{3}{320\pi^3C}}N^{1/6}\sim AN^{1/6}.
\end{equation}

\subsection{3D arrays with $d > 0.5\lambda_0$}

For $d>0.5\lambda_0$, the square of the sum over reciprocal lattices is separated in two terms. As in 2D, the term where $\gb = \gb'$ is exactly equivalent to the case when $d<0.5\lambda_0$ and so diverges as a power law with $N$. The term with $\gb \neq \gb'$ is a convergent integral and simply adds up to a constant $\tilde{B}$, which is the approximation of the slow varying function $\tilde{B}(d)$. Using the condition given in Eq.~\eqref{condition}, $d_{\text{critical}}$ is the solution to the following equation for large $N$ values
\begin{equation}
    \frac{3\lambda_0^3}{160\pi^3d^3}\left(\frac{N^{1/3}d}{C\lambda_0}+ \tilde{B}\right)-2=0,
\end{equation}
which is a third degree polynomial with a known real root that can be approximated to give the scaling of $d_{\text{critical}}$ with atom number as
\begin{equation}
    \dcrit \sim \bar{A}N^{1/6} + \bar{B}.
\end{equation}
where $\bar{A}, \bar{B}$ depend on the set of relevant reciprocal lattice vectors.

\end{document}